\documentclass[prb,superscriptaddress,floatfix,twocolumn,showpacs,amsfonts,amsmath,amssymb,final,superscriptaddress]{revtex4}

\usepackage{graphicx}
\usepackage{stmaryrd}
\usepackage{rotating}
\usepackage{wasysym}
\usepackage{amsmath}
\usepackage{amsfonts}
\usepackage{amssymb}
\usepackage[countmax]{subfloat}
\usepackage{dcolumn} 
\usepackage{bm} 
\usepackage{color}

\usepackage{float}
\usepackage{latexsym}

\begin{document}

\title{Spin Hall Effect Induced Spin Transfer Through an Insulator}


\author{Wei Chen}  

\affiliation{Institut f\"ur Theoretische Physik, ETH-Z\"urich, CH-8093 Z\"urich, Switzerland}

\author{Manfred Sigrist}

\affiliation{Institut f\"ur Theoretische Physik, ETH-Z\"urich, CH-8093 Z\"urich, Switzerland}

\author{Dirk Manske}

\affiliation{Max-Planck-Institut f\"{u}r Festk\"{o}rperforschung, Heisenbergstrasse 1, D-70569 Stuttgart, Germany}

\date{\rm\today}

\begin{abstract}

{ When charge current passes through a normal metal that exhibits spin Hall effect, spin accumulates at the edge of the sample in the transverse direction. We predict that this spin accumulation, or spin voltage, enables quantum tunneling of spin through an insulator or vacuum to reach a ferromagnet without transferring charge. In a normal metal/insulator/ferromagnetic insulator trilayer (such as Pt/oxide/YIG), the quantum tunneling explains the spin-transfer torque and spin pumping that exponentially decay with the thickness of the insulator. In a normal metal/insulator/ferromagnetic metal trilayer (such as Pt/oxide/Co), the spin transfer in general does not decay monotonically with the thickness of the insulator. Combining with the spin Hall magnetoresistance, this tunneling mechanism points to the possibility of a new type of tunneling spectroscopy that can probe the magnon density of states of a ferromagnetic insulator in an all-electrical and noninvasive manner. }

\end{abstract}






\maketitle

\section{Introduction}

A major issue in the field of spintronic research concerns the electrical control of magnetization dynamics. A particularly feasible scheme is to utilize the spin Hall effect (SHE)\cite{Dyakonov71,Hirsch99,Sinova15} in a normal metal (NM), where an applied charge current causes a spin accumulation, or spin voltage, at the transverse edge of the sample\cite{Zhang00,Takahashi06,Kato04}. When the edge is in conjunction with a ferromagnetic insulator (FMI) such as Y$_{3}$Fe$_{5}$O$_{12}$ (YIG) \cite{Kajiwara10}, or thin film ferromagnetic metal (FMM) such as Co \cite{Miron10,Miron11,Liu12,Liu12_2,Garello13}, the spin voltage induces a spin-transfer torque (STT) \cite{Berger96,Slonczewski96} on the magnetization, rendering an efficient mechanism for magnetization switching. In the reciprocal process known as spin pumping\cite{Tserkovnyak02,Zhang04}, a magnetization dynamics induced by, for instance, ferromagnetic resonance (FMR), injects a pure spin current into the NM\cite{Hahn13,Wei14,Weiler14}.

The microscopic origin of these fascinating phenomena has been linked to the quantum tunneling of spin without transferring charge\cite{Chen15}, which states that the spin voltage causes an injection of spin-polarized electrons towards the FMI or FMM thin film. The electrons are totally reflected back but have finite probability of flipping their spin after reflection, hence transfer angular momentum into the FMI or FMM. In this article, we further predict that even if the FMI or FMM is not in direct contact with the NM, but separated by an insulating oxide layer or vacuum, the spin-polarized electron can still tunnel through the separation to cause spin transfer. Such a tunneling process serves as the spintronic analog of field electron emission, in the sense that it transfers only spin but no charge through an insulating barrier (in contrast to a magnetic tunnel junction that transfers both charge and spin), and is purely a quantum effect that may be overlooked by diffusive approaches.

Based on a minimal model for the quantum tunneling of spins\cite{Chen15} that incorporates the Onsager relation between STT and spin pumping\cite{Maekawa12}, our theory well explains the spin pumping experiment in Pt/oxide/YIG performed by Du {\it et al.} that reveals a spin pumping spin current that exponentially decays with the oxide thickness, with a decay length related to the square root of the tunneling barrier\cite{Du13}. On the other hand, when this tunneling mechanism is applied to a NM/oxide/FMM trilayer (such as Pt/oxide/Co) with an FMM that is thinner than its spin relaxation length, we predict that the quantum interference may render a spin transfer that does not simply decay with the oxide thickness. Furthermore, since field electron emission is the basis of scanning tunneling microscopy (STM), we explore the possibility of a tunneling spectroscopy based on this quantum tunneling of spins. The result is a new type of tunneling spectroscopy that has direct access to the magnon excitation in an FMI, yet the measurement in reality may be very challenging.

The structure of the article is arranged in the following way. In Sec.~II, we give a general formalism for the quantum tunneling induced by the SHE in NM/oxide/FMI and NM/oxide/FMM, and calculate the dependence of the tunneling spin mixing conductance on generic material properties such as insulating gap, interface exchange coupling, and oxide thickness. Sec.~III formulates a tunneling spectroscopy for the NM/oxide/FMI and show how it is related to the magnetoresistance recently measured in these systems, and discuss the challenge of measuring the proposed differential conductance in reality. Sec.~IV summarizes the results. 






\section{Quantum tunneling of spin through an insulator or vacuum}

\begin{figure}[ht]
\begin{center}
\includegraphics[clip=true,width=0.99\columnwidth]{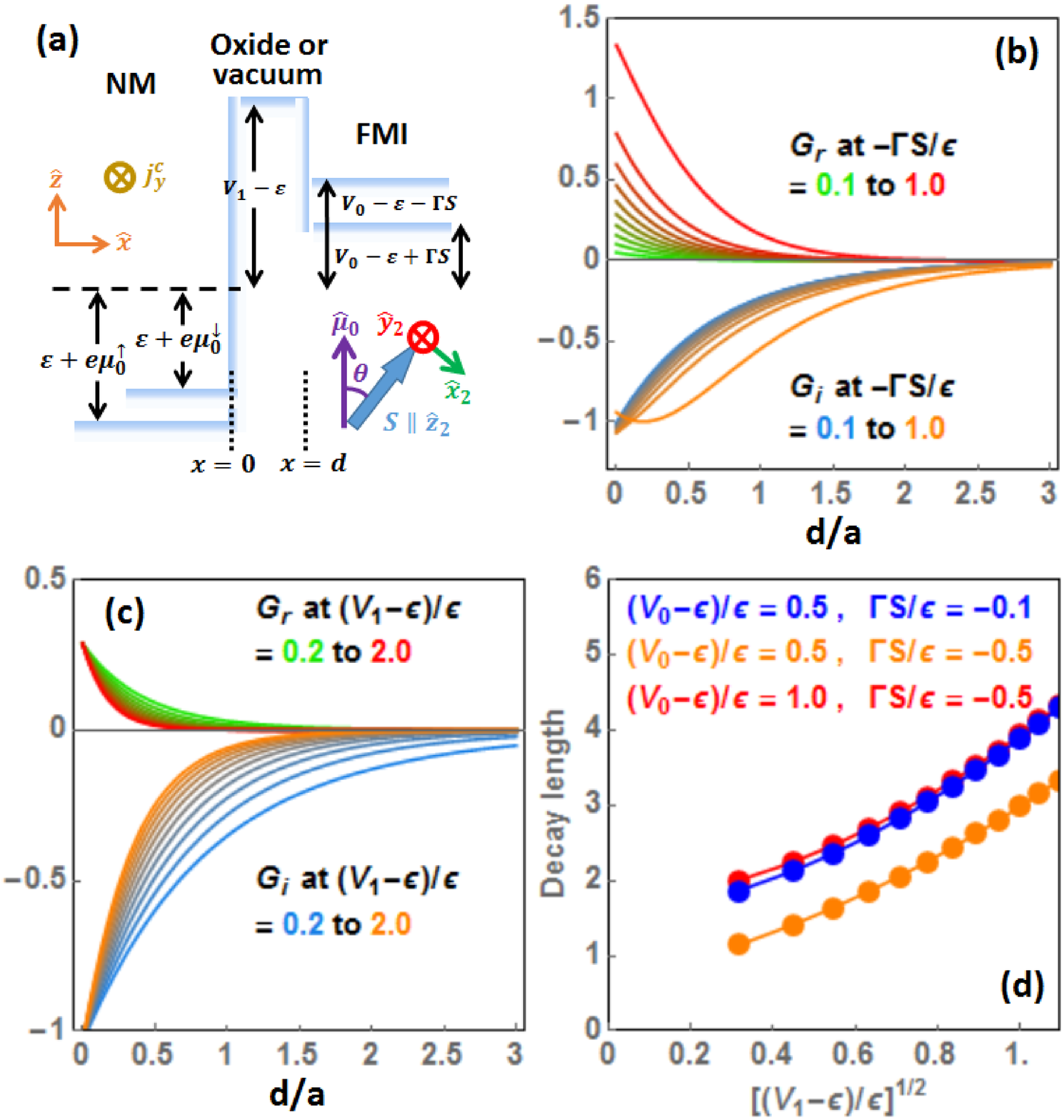}
\caption{(color online) (a) Schematics of the NM/oxide/FMI set up. (b) Spin mixing conductance $G_{r,i}$ as a function of the oxide thickness $d/a$ and interface $s-d$ coupling $-\Gamma S/\epsilon$, at FMI gap $(V_{0}-\epsilon)/\epsilon=1$ and oxide gap $(V_{1}-\epsilon)/\epsilon=0.5$, plotted in units of $e^{2}/\hbar a^{2}$ which is about $10^{14}\sim 10^{15}\Omega^{-1}$m$^{-2}$ depending on the Fermi wave length $a$. (c) $G_{r,i}$ as a function of the oxide thickness $d/a$ at different values of oxide gap  $(V_{1}-\epsilon)/\epsilon$, at FMI gap $(V_{0}-\epsilon)/\epsilon=1$ and $s-d$ coupling $-\Gamma S/\epsilon=0.5$. (d) The decay length of $G_{r}$ versus square root of the oxide gap or work function $\sqrt{(V_{1}-\epsilon)/\epsilon}$ at several parameters.  } 
\label{fig:NMoxideFMI_Gr_Gi}
\end{center}
\end{figure}

To demonstrate the spin voltage induced quantum tunneling through a separation layer, we adopt the formalism developed in the minimal model in Ref.~\onlinecite{Chen15}. Consider the NM/oxide/FMI trilayer shown in Fig.~\ref{fig:NMoxideFMI_Gr_Gi} (a) that contains three regions: (1) An NM at $-\infty<x<0$ described semiclassically by $H_{N}=p^{2}/2m-\mu_{x}^{\sigma}$, where $\mu_{x}^{\sigma}=\pm|{\boldsymbol\mu}_{x}|/2$ is the spin voltage of spin $\sigma=\left\{\uparrow,\downarrow\right\}$ at position $x$ caused by SHE\cite{Zhang00,Takahashi06,Kato04}. For an up spin incident from the left, the wave function at $x<0$ is
\begin{eqnarray}
\psi_{N}=\left(Ae^{ik_{0\uparrow}x}
+Be^{-ik_{0\uparrow}x}\right)\left(
\begin{array}{l}
1 \\
0
\end{array}
\right)
+Ce^{-ik_{0\downarrow}x}\left(
\begin{array}{l}
0 \\
1
\end{array}
\right),
\label{NM_wave_function}
\end{eqnarray}
where $k_{0\sigma}=\sqrt{2m\left(\epsilon+\mu_{0}^{\sigma}\right)}/\hbar$, and $\epsilon$ is the Fermi energy that serves as energy unit. We consider a charge current $j_{y}^{c}{\hat{\bf y}}$, so electrons moving in ${\hat{\bf x}}$ direction has ${\boldsymbol\sigma}\parallel{\hat{\bf z}}$ because of the SHE relation $k_{0\sigma}{\hat{\bf x}}\parallel{\boldsymbol\sigma}\times{\hat{\bf y}}$, and gives a positive spin voltage ${\boldsymbol\mu}_{0}\parallel{\hat{\bf z}}$ at the interface such that $k_{0\uparrow}>k_{0\downarrow}$. (2) Vacuum or nonmagnetic oxide in the region $0<x<d$ described by $H_{O}=p^{2}/2m+V_{1}$ with $V_{1}-\epsilon>0$ the work function or insulating gap, whose wave function is
\begin{eqnarray}
\psi_{O}=\left(De^{-\lambda x}+Ee^{\lambda x}\right)\left(
\begin{array}{l}
1 \\
0
\end{array}
\right)+\left(Fe^{-\lambda x}+Ge^{\lambda x}\right)\left(
\begin{array}{l}
0 \\
1
\end{array}
\right),
\label{oxide_wave_function}
\end{eqnarray}
where $\lambda=\sqrt{2m(V_{1}-\epsilon)}/\hbar$. (3) The FMI occupying $x>d$ described by 
$H_{FI}=p^{2}/2m+V_{0}+\Gamma{\bf S}\cdot{\boldsymbol \sigma}$, 
where $V_{0}>\epsilon$ is the potential step. The ${\bf S}\cdot{\boldsymbol \sigma}$ term describes the $s-d$ hybridization of the conduction electron spin and the magnetization ${\bf S}=S(\sin\theta\cos\varphi,\sin\theta\sin\varphi,\cos\theta)$. The evanescent wave function in the FMI is 
\begin{eqnarray}
\psi_{FI}&=&He^{-q_{+}x}
\left(
\begin{array}{l}
e^{-i\varphi/2}\cos\frac{\theta}{2} \\
e^{i\varphi/2}\sin\frac{\theta}{2} 
\end{array}
\right)
\nonumber \\
&&+Ie^{-q_{-}x}
\left(
\begin{array}{l}
-e^{-i\varphi/2}\sin\frac{\theta}{2} \\
e^{i\varphi/2}\cos\frac{\theta}{2} 
\end{array}
\right)\;,
\label{wave_fn_region_2}
\end{eqnarray}
where $q_{\pm}=\sqrt{2m\left(V_{0}\pm\Gamma S-\epsilon\right)}/\hbar>0$.

The amplitudes $A\sim E$ are determined by matching the wave function and its derivative at $x=0$ and $x=d$, leaving only one free variable that is attributed to the Fermi surface-averaged spin density at the interface $|A|^{2}=N_{F}|{\boldsymbol\mu}_{0}|/a^{3}$, where $N_{F}$ is the density of states (DOS) per $a^{3}$ at the Fermi surface, and $a=2\pi/k_{F}=h/\sqrt{2m\epsilon}$ is the Fermi wave length that serves as unit length. It is convenient\cite{Berger96} to introduce the frame $(x_{2},y_{2},z_{2})$ defined in Fig.~\ref{fig:NMoxideFMI_Gr_Gi} (a), where ${\hat{\bf z}_{2}}\parallel{\bf S}$, ${\hat{\bf y}_{2}}={\hat{\boldsymbol\mu}_{0}}\times{\hat{\bf S}}/\sin\theta$, and ${\hat{\bf x}_{2}}=-{\hat{\bf S}}\times\left({\hat{\boldsymbol\mu}_{0}}\times{\hat{\bf S}}\right)/\sin\theta$. The spinors in Eq.~(\ref{wave_fn_region_2}) are simply $(1\;0)^{T}$ and $(0\; 1)^{T}$ in this frame. The conduction electron spin tunneled into the FMI is $\langle{\boldsymbol\sigma}\rangle=\langle\psi_{FI}|{\boldsymbol\sigma}|\psi_{FI}\rangle$, whose components are
\begin{eqnarray}
\langle\sigma^{x_{2},y_{2}}\rangle&=&-16\frac{|A|^{2}}{|\gamma_{\theta}|^{2}}\sin\theta e^{-\left(q_{+}+q_{-}\right)(x-d)}
\nonumber \\
&&\times\left({\rm Re},{\rm Im}\right)\left(W_{\downarrow-}^{\ast}W_{\downarrow+}\right)\;,
\label{SxSy_in_CMI}
\end{eqnarray}
where $W_{\sigma\pm}$ and $\gamma_{\theta}$ are defined in the Appendix A. The total spin per unit area $a^{2}$ is denoted by $\langle\overline{{\boldsymbol\sigma}}\rangle=a^{2}\int_{0}^{\infty}dx\langle{\boldsymbol\sigma}\rangle$.
The magnetization within the range of $\langle\overline{\boldsymbol\sigma}\rangle$, about $2\pi/\left(q_{+}+q_{-}\right)\sim a$, is treated as a macrospin ${\bf S}$. From the Landau-Lifshitz (LL) dynamics, the $s-d$ coupling $H_{sd}=\Gamma{\boldsymbol\sigma}\cdot{\bf S}$ renders the STT\cite{Berger96}, whose response in the damping-like and field-like direction define the spin mixing conductance\cite{Maekawa12} $G_{r}$ and $G_{i}$, respectively,
\begin{eqnarray}
{\boldsymbol\tau}&=&\frac{\Gamma}{\hbar}\langle\overline{{\boldsymbol\sigma}}\rangle\times{\bf S}=\frac{\Gamma}{\hbar}S\langle\overline{\sigma}^{y_{2}}\rangle{\hat{\bf x}_{2}}-\frac{\Gamma}{\hbar}S\langle\overline{\sigma}^{x_{2}}\rangle{\hat{\bf y}_{2}}
\nonumber \\
&=&\frac{\Gamma Sa^{2}N_{F}}{\hbar}\left[G_{r}{\hat{\bf S}}\times\left({\hat{\bf S}}\times{\boldsymbol\mu}_{0}\right)+G_{i}{\hat{\bf S}}\times{\boldsymbol\mu}_{0}\right]
\nonumber \\
G_{r,i}&=&\int_{d}^{\infty}\frac{\langle\sigma^{y_{2},x_{2}}\rangle}{N_{F}|{\boldsymbol\mu}_{0}|\sin\theta}dx
=-16\frac{\left({\rm Im},{\rm Re}\right)\left(W_{\downarrow-}^{\ast}W_{\downarrow+}\right)}{a^{3}|\gamma_{\theta}|^{2}\left(q_{+}+q_{-}\right)}\;,
\nonumber \\
\label{effective_EOM}
\end{eqnarray}
after substituting Eq.~(\ref{SxSy_in_CMI}) and $|A|^{2}=N_{F}|{\boldsymbol\mu}_{0}|/a^{3}$. Alternatively, from angular momentum conservation $a^{2}({\boldsymbol j}_{0}-{\boldsymbol j}_{\infty})=a^{2}{\boldsymbol j}_{0}=a^{2}{\boldsymbol j}_{d}={\boldsymbol\tau}$ where ${\boldsymbol j}_{x}$ is the spin current at position $x$, one obtains the same $G_{r,i}$\cite{Chen15}. The phenomenon of spin pumping has also been demonstrated in this set up\cite{Du13}, whose mechanism follows that discussed in Ref.~\onlinecite{Chen15} and satisfies Onsager relation\cite{Maekawa12}. In both STT and spin pumping, the spin relaxation in the FMI plays a relatively minor role\cite{Chen15}.

For most parameters, both $G_{r}$ and $G_{i}$ monotonically decay with the separation thickness $d$, consistent with the spin pumping experiment in Pt/oxide/YIG\cite{Du13}. However, when the exchange coupling is comparable to the insulating gap of the FMI, $|\Gamma S|\approx(V_{0}-\epsilon)/\epsilon$, we found that the damping-like component $G_{i}$ displays a slight enhancement at small $d$ ($\approx 0.2a\approx 0.08$nm for Pt), as shown by the orange line in Fig.~\ref{fig:NMoxideFMI_Gr_Gi} (b). Thus the interplay between the exchange coupling and the insulating gaps can lead to unconventional tunneling behaviors in certain parameter ranges. In most of the parameter regime, the $G_{r,i}$ versus oxide or vacuum thickness $d$ fits well with an exponentially decay form, with a decay length that decreases with increasing oxide gap or vacuum work function, as shown in Fig.~\ref{fig:NMoxideFMI_Gr_Gi} (c), in accordance with that found experimentally\cite{Du13}. A systematic investigation of the decay length of $G_{r}$ (proportional to d.c. component of spin pumping spin current) versus square root of the oxide gap $\sqrt{(V_{1}-\epsilon)/\epsilon}$, as suggested experimentally\cite{Du13}, reveals a behavior that sensitively depends on other parameters in this minimal model, such as the FMI gap and $s-d$ coupling, as shown in Fig.~\ref{fig:NMoxideFMI_Gr_Gi} (d). Although in large $\sqrt{(V_{1}-\epsilon)/\epsilon}$ the decay length seems rather linear to $\sqrt{(V_{1}-\epsilon)/\epsilon}$, it in general does not extract to zero in most of the parameter regimes, unlike that assumed experimentally\cite{Du13}.



\begin{figure}[ht]
\begin{center}
\includegraphics[clip=true,width=0.99\columnwidth]{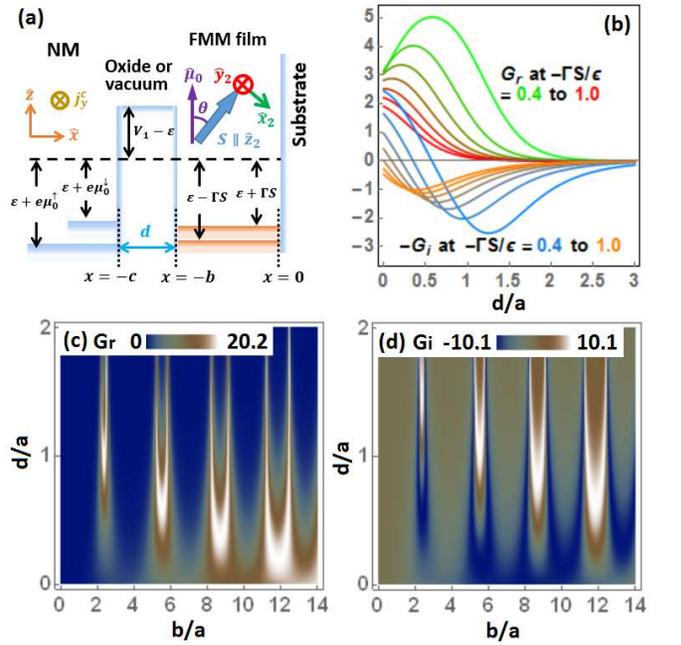}
\caption{(color online) (a) Schematics of the NM/oxide/FMM/substrate set up. (b) Spin mixing conductance $G_{r}$ and $G_{i}$ versus oxide or vacuum thickness $d$, where $a$ is the Fermi wave length, at FMM thickness $b=2a$ and several values of $s-d$ coupling $-\Gamma S/\epsilon$. (c) and (d) shows $G_{r}$ and $G_{i}$ versus FMM thickness $b$ and oxide or vacuum thickness $d$, at $s-d$ coupling $-\Gamma S/\epsilon=0.1$, plotted in units of $e^{2}/\hbar a^{2}$ where $a$ is the Fermi wave length. The oxide gap or work function in these plots are fixed at $(V_{1}-\epsilon)/\epsilon=1$.  } 
\label{fig:NMoxideFMMsubstrate_Gr_Gi}
\end{center}
\end{figure}

The same analysis is applicable to the NM/oxide/FMM/substrate multilayer if the FMM is thinner than its spin relaxation length. In this case, since the insulating gap of the nonmagnetic substrate is not crucial to spin transport\cite{Chen15}, we set it to be infinite for simplicity, such that the wave function vanishes inside the substrate. Since the oxide separates NM and FMM, we consider the situation that in the NM a charge current $j_{y}^{c}{\hat{\bf y}}$ flows but not in the FMM, such that the spin-orbit torque\cite{Manchon08,Haney10,Pesin12,Kim12,Haney13,Wang12,Gambardella11,Kurebayashi14} does not arise, and the spin torque comes entirely from the SHE in the NM. The wave function in the NM and in the oxide are described by Eqs.~(\ref{NM_wave_function}) and (\ref{oxide_wave_function}), with the interface positions defined in Fig.~\ref{fig:NMoxideFMMsubstrate_Gr_Gi}(a). The wave function in the FMM described by $H_{FM}=p^{2}/2m+\Gamma{\bf S}\cdot{\boldsymbol \sigma}$ is 
\begin{eqnarray}
\psi_{FM}&=&2iH\sin\left(k_{+}x\right)
\left(
\begin{array}{l}
e^{-i\varphi/2}\cos\frac{\theta}{2} \\
e^{i\varphi/2}\sin\frac{\theta}{2} 
\end{array}
\right)
\nonumber \\
&&+2iI\sin\left(k_{-}x\right)
\left(
\begin{array}{l}
-e^{-i\varphi/2}\sin\frac{\theta}{2} \\
e^{i\varphi/2}\cos\frac{\theta}{2} 
\end{array}
\right)\;,
\label{FMM_wave_function}
\end{eqnarray}
where $k_{\pm}=\sqrt{2m(\epsilon\mp\Gamma S)}/\hbar$. The spin expectation value in the FMM is
\begin{eqnarray}
\langle\sigma^{x_{2},y_{2}}\rangle&=&-64\frac{|A|^{2}}{|\gamma_{\theta}^{\prime}|^{2}}\sin\theta \sin k_{+}x\sin k_{-}x
\nonumber \\
&&\times\left({\rm Re},{\rm Im}\right)\left(U_{\downarrow-}^{\ast}U_{\downarrow+}\right)\;.
\label{SxSy_in_FMM}
\end{eqnarray}
where $U_{\sigma\pm}$ and $\gamma_{\theta}^{\prime}$ are defined in the Appendix A. Using Eq.~(\ref{effective_EOM}), the spin mixing conductance is 
\begin{eqnarray}
&&G_{r,i}=-\frac{64}{a^{3}|\gamma_{\theta}^{\prime}|^{2}}\left({\rm Im},{\rm Re}\right)\left\{U_{\downarrow-}^{\ast}U_{\downarrow+}\right.
\nonumber \\
&&\left.\times\frac{k_{-}\sin k_{+}b\cos k_{-}b-k_{+}\sin k_{-}b\cos k_{+}b}{k_{+}^{2}-k_{-}^{2}}\right\}\;.
\end{eqnarray}

Figure \ref{fig:NMoxideFMMsubstrate_Gr_Gi} (b) shows $G_{r}$ and $G_{i}$ versus thickness of the oxide or vacuum $d$. Remarkably, $G_{r}$ in general does not monotonically decrease with $d$ but may show significant enhancement (up to more than $50\%$) at small $d$, and $G_{i}$ can change sign with increasing $d$. This very unconventional tunneling behavior implies that, surprisingly, inserting an insulating oxide of atomic layer thickness may enhance $G_{r}$ and hence the efficiency of magnetization switching in FMM thin films\cite{Miron10,Miron11,Liu12,Liu12_2,Garello13}. Figure \ref{fig:NMoxideFMMsubstrate_Gr_Gi} (c) and (d) show $G_{r}$ and $G_{i}$ as functions of FMM thickness $b$ and oxide thickness $d$, where this nonmonotonic dependence on $d$ can be seen in many regions of parameter space. In addition, certain periodicity with respect to the FMM thickness $b$ is evident, a result expected from the quantum interference effect when the spin travels inside the FMM\cite{Chen15}. In Appendix A, we further demonstrate that the presence of the vacuum or oxide layer, despite being nonmagnetic, influences the quantum interference pattern of $G_{r,i}$.

\section{A novel tunneling spectroscopy that measures magnon DOS}

As field electron emission is the basis of tunneling spectroscopy, below we discuss the possibility of a new type of tunneling spectroscopy based on this quantum tunneling of spins without transferring charge. Consider the NM/vacuum/FMI in Fig.~\ref{fig:NMoxideFMI_Gr_Gi} (a). We aim at calculating the differential conductance defined from the spin current and spin voltage. Under the approximation that only the magnetization at the interface atomic layer denoted by ${\bf S}_{0}$ experiences the STT since electrons only penetrate the FMI over a very short distance, the STT is equivalent to the torque caused by applying an effective magnetic field on ${\bf S}_{0}$. Thus the FMI under the influence of the STT is described by 
\begin{eqnarray}
H_{FM}=H_{FM}^{0}+H_{int}=H_{FM}^{0}+{\bf B}_{eff}\cdot{\bf S}_{0}\;,
\label{H_FM}
\end{eqnarray}
where $H_{FM}^{0}$ describes the magnons. The effective magnetic field ${\bf B}_{eff}$ caused by STT originates from the total spin $\langle{\overline{\boldsymbol\sigma}}\rangle$ tunneled into the FMI 
\begin{eqnarray}
{\bf B}_{eff}=-\tilde{G}_{r}{\hat{\bf S}}_{0}\times{\boldsymbol\mu}_{0}-\tilde{G}_{i}{\boldsymbol\mu}_{0}\;,
\label{B_eff}
\end{eqnarray}
where $\tilde{G}_{r,i}=\left(\Gamma S a^{2}\right)G_{r,i}$, since ${\boldsymbol\tau}=d{\bf S}_{0}/dt=i\left[{\bf B}_{eff}\cdot{\bf S}_{0},{\bf S}_{0}\right]$ gives the correct LL dynamics described by Eq.~(\ref{effective_EOM}) with the replacement ${\bf S}\rightarrow{\bf S}_{0}$ (hereafter $\hbar=1$). Note that in Eqs.~(\ref{H_FM}) and (\ref{B_eff}), the unit vector ${\hat{\bf S}}_{0}=(\sin\theta\cos\varphi,\sin\theta\sin\varphi,\cos\theta)$ denotes the direction of the interface magnetization, while ${\bf S}_{0}=(S_{0}^{x},S_{0}^{y},S_{0}^{z})$ is the operator of interface magnetization expressed in the $(x,y,z)$ frame, and the spin accumulation ${\boldsymbol\mu}_{0}=c_{0\alpha}^{\dag}{\boldsymbol\sigma}_{\alpha\beta}c_{0\beta}$ is expressed in terms of electron operators at the interface.

The ${\bf B}_{eff}$, which contains both field-like and damping-like components, causes a torque that can be calculated within linear response theory\cite{Kajiwara10,Takahashi10}. We are interested in $z$ component of the torque as it is related to the longitudinal spin Hall magnetoresistance (SMR) \cite{Nakayama13,Chen13,Althammer13,Avci15} as addressed below. After Fourier transform and a gauge transformation, as demonstrated in Appendix B, the $z$-component of the torque operator and the relevant terms containing $S^{+}$ or $S^{-}$ in $H_{int}$ are 
\begin{eqnarray}
L(t)&=&\sum_{kk^{\prime}q}S_{q}^{-}(t)c_{k^{\prime}\uparrow}^{\dag}(t)c_{k\downarrow}(t)\;,
\nonumber \\
M(t)&=&\sum_{kk^{\prime}q}S_{q}^{-}(t)
\left(c_{k^{\prime}\uparrow}^{\dag}(t)c_{k\uparrow}(t)-c_{k^{\prime}\downarrow}^{\dag}(t)c_{k\downarrow}(t)\right)\;,
\nonumber \\
H_{int}^{rel}&=&iJ_{1}L^{\dag}(0)-iJ_{1}^{\ast}L(0)
+iJ_{2}\left[M^{\dag}(0)-M(0)\right]\;,
\nonumber \\
\hat{\tau}^{z}&=&J_{1}L^{\dag}(0)+J_{1}^{\ast}L(0)
+J_{2}\left[M^{\dag}(0)+M(0)\right]\;,
\label{operators_in_linear_response}
\end{eqnarray}
where $J_{1}=\left(\tilde{G}_{r}\cos\theta+i\tilde{G}_{i}\right)/\sqrt{N_{N}N_{I}}$ and $J_{2}=-\left(\tilde{G}_{r}\sin\theta\right)/2\sqrt{N_{N}N_{I}}$ are the effective coupling between NM and FMI, with $N_{N}$ and $N_{I}$ denoting the number of lattice sites for NM and FMI along ${\hat{\bf x}}$ direction, respectively. The spin conserved $L(t)$ and spin nonconserved $M(t)$ originate from the field-like and damping-like component of Eq.~(\ref{B_eff}), respectively.

The linear response theory\cite{Kajiwara10,Takahashi10}
\begin{eqnarray}
\tau^{z}=-i\int_{-\infty}^{t}dt^{\prime}\langle\left[{\hat\tau}^{z}(t),H_{int}^{rel}(t^{\prime})\right]\rangle\;,
\end{eqnarray}
with ${\hat\tau}^{z}(t)=e^{iH^{\prime}t}\hat{\tau}^{z}e^{-iH^{\prime}t}$ and $H_{int}^{rel}(t^{\prime})=e^{iH^{\prime}t^{\prime}}H_{int}^{rel}e^{-iH^{\prime}t^{\prime}}$, where the total Hamiltonian $H^{\prime}=H_{\uparrow}+H_{\downarrow}+H_{FM}^{0}$ describes the spin up and down electrons in the NM and magnons in the FMI, leads to
\begin{eqnarray}
\tau^{z}=-2|J_{1}|^{2}{\rm Im}\left[U_{ret}^{L}(\mu_{0})\right]-2|J_{2}|^{2}{\rm Im}\left[U_{ret}^{M}(0)\right]\;,
\label{tau_z_U_U}
\end{eqnarray}
as shown in Appendix B. The retarded response functions are calculated in the Matsubara version 
\begin{eqnarray}
U^{L}(i\omega)=-\int_{0}^{\beta}d\tau e^{i\omega\tau}\langle T_{\tau}L(\tau)L^{\dag}(\tau)\rangle
\end{eqnarray}
and then using analytical continuation $i\omega\rightarrow\mu_{0}+i\delta$, and similarly for $U^{M}(i\omega)$ using $i\omega\rightarrow 0+i\delta$. Only the first term in Eq.~(\ref{tau_z_U_U}) is nonzero, yielding (see Appendix B)
\begin{widetext}
\begin{eqnarray}
\tau^{z}&=&4\pi \langle S^{z}\rangle|J_{1}|^{2}\int d\xi\int d\Omega \;N_{\uparrow}\left(\xi\right)N_{\downarrow}\left(\xi+\mu_{0}+\Omega\right)N_{M}(\Omega)
\nonumber \\
&&\times\left[n_{F}(\xi+\mu_{0}+\Omega)-n_{F}(\xi)\right]\left[n_{B}(-\Omega-\mu_{0})-n_{B}(-\Omega)\right]=a^{2}j_{0}^{z}\;,
\end{eqnarray}
\end{widetext}
where we have converted momentum sums into energy integrals, and used angular momentum conservation to identify the torque with the spin current in the NM at the interface $j_{0}^{z}$. The $N_{\uparrow}$ and $N_{\downarrow}$ are the DOS of spin up and down electrons that are assumed to be constant within the range of measurement, and $N_{M}(\Omega)$ is the DOS of magnons at $\Omega$. The $n_{F}$ and $n_{B}$ denote the Fermi and Bose distribution function, and $\langle S^{z}\rangle$ is the ground state magnetization of the FMI in the linear spin wave theory. As demonstrated in Appendix B, the derivative of the spin current with respect to spin voltage is linear in temperature $T$ and proportional to the magnon DOS at $-\mu_{0}$
\begin{eqnarray}
\frac{dj_{0}^{z}}{d\mu_{0}}\propto k_{B}T\langle S^{z}\rangle|J_{1}|^{2} N_{\uparrow}N_{\downarrow}N_{M}(-\mu_{0})
\propto\frac{d\Delta\rho_{1}}{d j_{c}}\;,\;
\label{differential_conductance}
\end{eqnarray}
where we have used the fact that $j_{0}^{z}$, after converted back to a charge current via inverse spin Hall effect (ISHE), is proportional to the SMR in the longitudinal direction $\Delta\rho_{1}$, and $\mu_{0}$ is proportional to the input charge current $j_{c}$ \cite{Chen13}. Equation (\ref{differential_conductance}) indicates a new type of tunneling spectroscopy that can directly measure the magnon DOS of the FMI at a specific energy $-\mu_{0}$, achieved by taking the derivative of the longitudinal SMR in the NM with respect to the input charge current.

The analysis from Eq.~(\ref{H_FM}) to (\ref{differential_conductance}) is also valid if the oxide or vacuum is absent ($d=0$ in Eqs.(\ref{SxSy_in_CMI}) to (\ref{effective_EOM})), i.e., an NM/FMI bilayer in which SMR has been intensively investigated\cite{Nakayama13,Althammer13,Chen13}, although the probe (NM) is permanently attached to the sample (FMI) in this situation. These SMR experiments typically operate at charge current density no more than $j_{c}\sim 10^{8}$A/cm$^2$ at room temperature due to Joule heating \cite{Nakayama13,Althammer13,Avci15}. Assuming a typical NM thickness $\sim 10$nm and a spin diffusion length of the same order, and the conductivity $\sigma\sim 10^{7}$S/m and spin Hall angle $\theta_{H}\sim 0.1$ for Pt\cite{Sinova15}, the spin voltage produced at $j_{c}\sim 10^{8}$A/cm$^2$ is about $\mu_{0}\sim$0.1meV, and so is the maximal energy at which magnon DOS can be probed according to Eq.~(\ref{differential_conductance}). Compared to the whole magnon band width that is typically about $10\sim 100$meV in solids, the proposed tunneling spectroscopy therefore measures the magnons at very low energy that are generally more coherent and  well described by the linear spin wave theory\cite{Kajiwara10,Chen15_2,Takahashi10,Barker16}. Comparing to other methods that measure magnon DOS, this probing energy range is few orders higher than that uses nitrogen-vacancy (NV) center in diamonds ($\sim\mu$eV)\cite{vanderSar15}, and approaching the probing range of the Brillouin light scattering ($\sim$meV)\cite{Sebastian15}, while having the advantage of being an all-electrical measurement that requires no additional field or light source.

One can estimate the change of spin current $\Delta j_{0}^{z}$ due to excitation of magnons from Eq.~(\ref{differential_conductance}). The prefactor of the proportionality in Eq.~(\ref{differential_conductance}) is of the order of $4\pi/a^{2}\hbar$ times unity for an NM/FMI bilayer. Assuming the charge current in the NM is increased from zero to a maximal $j_{c}\sim 10^{8}$A/cm$^{2}$, which gives a change of spin voltage $\Delta\mu_{0}\sim 0.1$meV as discussed in the previous paragraph. Using $d j_{0}^{z}/d\mu_{0}\sim\Delta j_{0}^{z}/\Delta\mu_{0}$, and typical material parameters for the DOS $N_{\uparrow}\sim N_{\downarrow}\sim N_{M}\sim 1/$eV, layer thickness $\sqrt{N_{N}N_{I}}\sim 10$, exchange coupling $J_{1}\sim \Gamma Sa^{2}G_{r,i}/\sqrt{N_{N}N_{I}}\sim 0.01$eV, temperature $k_{B}T\sim 0.1$eV, and lattice constant $a\sim$ nm, one obtains the change of spin current (particle flux) $\Delta j_{0}^{z}\sim 10^{2}$A/cm$^{2}$. This should be compared with the SHE spin current at this maximal charge current, which is of the order of $j_{0,SHE}^{z}\sim\theta_{H}j_{c}\sim 10^{7}$A/cm$^{2}$ assuming the thickness of the NM is close to its spin diffusion length\cite{Chen13}. Therefore the magnon excitation gives a very small correction to the SHE spin current $\Delta j_{0}^{z}/j_{0,SHE}^{z}\sim 10^{-5}$, and so is the correction to SMR\cite{Chen13} at this maximal charge current $j_{c}\sim 10^{8}$A/cm$^{2}$, which can be very challenging to measure. Finally, we remark that in principle, an STM based on this quantum tunneling of spin is also possible by fabricating the NM into a nanometer size tip, although measuring the SMR caused by tunneling through such a small tip will obviously be very difficult.



\section{Conclusions}

In summary, we predict that the spin voltage caused by SHE can induce quantum tunneling of spin through a thin insulator or vacuum, realizing the spintronic analog of field electron emission. In the NM/oxide/FMI trilayer, this tunneling process yields a STT and spin pumping that in most of the parameter regime decays monotonically with the oxide thickness, in good agreement with the spin pumping experiment in Pt/oxide/YIG\cite{Du13}. In the NM/oxide/FMM, the quantum tunneling yields a spin mixing conductance that in general does not monotonically decay with thickness of the insulator or vacuum. Consequently, inserting an ultrathin insulator between NM and FMM may surprisingly improve the performance of the magnetization switching caused by SHE. For the NM/oxide/FMI case, a new type of tunneling spectroscopy is revealed based on this quantum tunneling of spins, which can directly probe the magnon DOS of the FMI, and has the advantage of being an all-electrical measurement that requires no external field or light source. Combining with SMR measurements, such a tunneling spectroscopy can be practically realized by taking the derivative of longitudinal SMR with respect to the input charge current, as described by Eq.~(\ref{differential_conductance}), although the smallness of the spin voltage may render such measurement rather challenging in reality.

We thank P. Gambardella for inspiring this project, and Y.-H. Liu, S. Ok, F. Casola, R. Wiesendanger, H. Schulthei\ss, and J. Mendil for fruitful discussions. W. C. and M. S. are grateful for the financial support through a research grant of the Swiss National Science Foundation.

\appendix

\section{Detail of spin-transfer torque calculation }


By matching the wave function at the $x=0$ and $x=a$ interfaces, we obtain the scattering coefficients in the NM/vacuum/FMI trilayer. Introducing  
\begin{eqnarray}
W_{\sigma\alpha}&=&\sum_{\beta=\pm}\left(1-\frac{\beta\lambda}{ik_{0\sigma}}\right)\left(1+\frac{q_{\alpha}}{\beta\lambda}\right)
e^{\beta\lambda d}\;,
\nonumber \\
\gamma_{\theta}&=&\sin^{2}\frac{\theta}{2}W_{\downarrow+}W_{\uparrow-}
+\cos^{2}\frac{\theta}{2}W_{\downarrow-}W_{\uparrow+}\;.
\nonumber \\
Z_{1\theta}&=&\frac{e^{\lambda a}}{\gamma_{\theta}}\left[\left(1+\frac{q_{+}}{\lambda}\right)\cos^{2}\frac{\theta}{2}W_{\downarrow-}\right.
\nonumber \\
&&\left.+\left(1+\frac{q_{-}}{\lambda}\right)\sin^{2}\frac{\theta}{2}W_{\downarrow+}\right]\;,
\nonumber \\
Z_{2\theta}&=&\frac{e^{-\lambda a}}{\gamma_{\theta}}\left[\left(1-\frac{q_{+}}{\lambda}\right)\cos^{2}\frac{\theta}{2}W_{\downarrow-}\right.
\nonumber \\
&&\left.+\left(1-\frac{q_{-}}{\lambda}\right)\sin^{2}\frac{\theta}{2}W_{\downarrow+}\right]\;,
\nonumber \\
Z_{3\theta}&=&\frac{e^{\lambda a}}{\gamma_{\theta}}\sin\frac{\theta}{2}\cos\frac{\theta}{2}\left[\left(1+\frac{q_{+}}{\lambda}\right)W_{\downarrow-}\right.
\nonumber \\
&&\left.-\left(1+\frac{q_{-}}{\lambda}\right)W_{\downarrow+}\right]\;,
\nonumber \\
Z_{4\theta}&=&\frac{e^{-\lambda a}}{\gamma_{\theta}}\sin\frac{\theta}{2}\cos\frac{\theta}{2}\left[\left(1-\frac{q_{+}}{\lambda}\right)W_{\downarrow-}\right.
\nonumber \\
&&\left.-\left(1-\frac{q_{-}}{\lambda}\right)W_{\downarrow+}\right]\;,
\end{eqnarray}
the scattering coefficients are written as
\begin{eqnarray}
B&=&A\left(1+\frac{\lambda}{ik_{0\uparrow}}\right)Z_{1\theta}
+A\left(1-\frac{\lambda}{ik_{0\uparrow}}\right)Z_{2\theta}\;,
\nonumber \\
C&=&Ae^{i\varphi}\left(1+\frac{\lambda}{ik_{0\downarrow}}\right)Z_{3\theta}
+Ae^{i\varphi}\left(1-\frac{\lambda}{ik_{0\downarrow}}\right)Z_{4\theta}\;,
\nonumber \\
D&=&2AZ_{1\theta}\;,\;
E=2AZ_{2\theta}\;,\;
\nonumber \\
F&=&2Ae^{i\varphi}Z_{3\theta}\;,\;
G=2Ae^{i\varphi}Z_{4\theta}\;,
\nonumber \\
H&=&4A\frac{e^{q_{+}a}}{\gamma_{\theta}}e^{i\varphi/2}\cos\frac{\theta}{2}W_{\downarrow-}\;,\;
\nonumber \\
I&=&-4A\frac{e^{q_{-}a}}{\gamma_{\theta}}e^{i\varphi/2}\sin\frac{\theta}{2}W_{\downarrow+}\;.
\end{eqnarray}
Since the particle flux is zero $k_{0\uparrow}|A|^{2}-k_{0\uparrow}|B|^{2}-k_{0\downarrow}|C|^{2}=0$, there is no charge current in this problem but only spin current, making it clear that the formalism describes the spin injection caused by the pure spin current in SHE.


The scattering coefficients for the NM/vacuum/FMM/substrate multilayer are obtained similarly from matching the wave function at the interfaces. Defining 
\begin{eqnarray}
\eta_{\alpha\beta}&=&-2i\left(\sin k_{\alpha}b+\beta\frac{k_{\alpha}}{\lambda}\cos k_{\alpha}b\right)\;,
\nonumber \\
U_{\sigma\alpha}&=&\sum_{\beta=\pm}e^{\beta\lambda(c-b)}\left(1-\frac{\beta\lambda}{ik_{\sigma}}\right)\eta_{\alpha\beta}\;,
\nonumber \\
\gamma_{\theta}^{\prime}&=&\sin^{2}\frac{\theta}{2}U_{\downarrow+}U_{\uparrow-}
+\cos^{2}\frac{\theta}{2}U_{\downarrow-}U_{\uparrow+}\;,
\nonumber \\
Y_{1\theta}&=&\eta_{++}\cos^{2}\frac{\theta}{2}U_{\downarrow-}+\eta_{-+}\sin^{2}\frac{\theta}{2}U_{\downarrow+}\;,
\nonumber \\
Y_{2\theta}&=&\eta_{+-}\cos^{2}\frac{\theta}{2}U_{\downarrow-}+\eta_{--}\sin^{2}\frac{\theta}{2}U_{\downarrow+}\;,
\nonumber \\
Y_{3\theta}&=&e^{i\varphi}\sin\frac{\theta}{2}\cos\frac{\theta}{2}\left(\eta_{++}U_{\downarrow-}-\eta_{-+}U_{\downarrow+}\right)\;,
\nonumber \\
Y_{4\theta}&=&e^{i\varphi}\sin\frac{\theta}{2}\cos\frac{\theta}{2}\left(\eta_{+-}U_{\downarrow-}-\eta_{--}U_{\downarrow+}\right)\;,
\end{eqnarray} 
we write
\begin{eqnarray}
B&=&Ae^{-2ik_{0\uparrow}c}\frac{e^{\lambda(c-b)}}{\gamma_{\theta}^{\prime}}\left(1+\frac{\lambda}{ik_{0\uparrow}}\right)Y_{1\theta}
\nonumber \\
&&+Ae^{-2ik_{0\uparrow}c}\frac{e^{-\lambda(c-b)}}{\gamma_{\theta}^{\prime}}\left(1-\frac{\lambda}{ik_{0\uparrow}}\right)Y_{2\theta}\;,
\nonumber \\
C&=&Ae^{-i(k_{0\uparrow}+k_{0\downarrow})c}\frac{e^{\lambda(c-b)}}{\gamma_{\theta}^{\prime}}\left(1+\frac{\lambda}{ik_{0\downarrow}}\right)Y_{3\theta}
\nonumber \\
&&+Ae^{-i(k_{0\uparrow}+k_{0\downarrow})c}\frac{e^{-\lambda(c-b)}}{\gamma_{\theta}^{\prime}}\left(1-\frac{\lambda}{ik_{0\downarrow}}\right)Y_{4\theta}\;,
\nonumber \\
D&=&2Ae^{-ik_{0\uparrow}c}\frac{e^{-\lambda b}}{\gamma_{\theta}^{\prime}}Y_{1\theta}\;,\;
E=2Ae^{-ik_{0\uparrow}c}\frac{e^{\lambda b}}{\gamma_{\theta}^{\prime}}Y_{2\theta}\;,
\nonumber \\
F&=&2Ae^{-ik_{0\uparrow}c}\frac{e^{-\lambda b}}{\gamma_{\theta}^{\prime}}Y_{3\theta}\;,\;
G=2Ae^{-ik_{0\uparrow}c}\frac{e^{\lambda b}}{\gamma_{\theta}^{\prime}}
Y_{4\theta}\;,
\nonumber \\
H&=&4Ae^{-ik_{0\uparrow}c}\frac{e^{i\varphi/2}}{\gamma_{\theta}^{\prime}}\cos\frac{\theta}{2}U_{\downarrow-}\;,\;
\nonumber \\
I&=&-4Ae^{-ik_{0\uparrow}c}\frac{e^{i\varphi/2}}{\gamma_{\theta}^{\prime}}\sin\frac{\theta}{2}U_{\downarrow+}\;.
\end{eqnarray}
The charge current in the NM is again zero since $k_{0\uparrow}|A|^{2}-k_{0\uparrow}|B|^{2}-k_{0\downarrow}|C|^{2}=0$.

\begin{widetext}
\begin{figure}[ht]
\begin{center}
\includegraphics[clip=true,width=1.99\columnwidth]{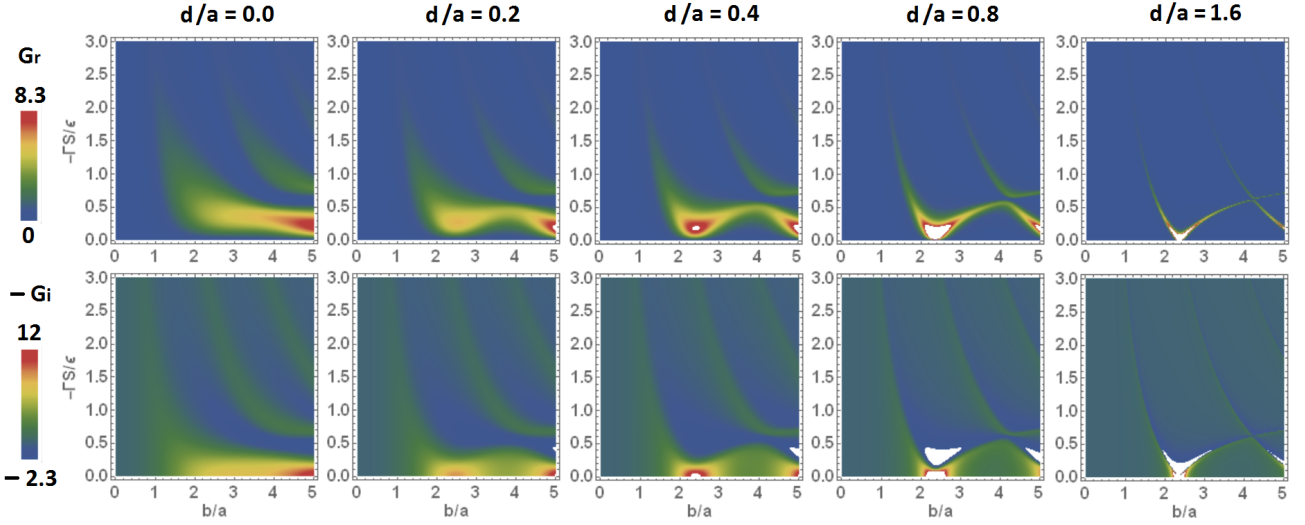}
\caption{(color online) The spin mixing conductance $G_{r}$ (upper figures) and $-G_{i}$ (lower figures) in NM/vacuum/FMM/substrate multilayer plotted as functions of FMM thickness $b$ and the $s-d$ coupling $-\Gamma S/\epsilon$, at different vacuum thickness $d$. The length unit $a$ is the Fermi wave length. The work function for tunneling through the vacuum is fixed at $(V_{1}-\epsilon)/\epsilon=1$.  } 
\label{fig:NMoxideFMMsubstrate_scanoxidethickness}
\end{center}
\end{figure}
\end{widetext}

Figure \ref{fig:NMoxideFMMsubstrate_scanoxidethickness} shows the spin mixing conductance $G_{r}$ and $-G_{i}$ as functions of FMM thickness $b$ and interface $s-d$ coupling $-\Gamma S/\epsilon$, for several values of vacuum thickness $d$ assuming a constant work function or oxide gap $(V_{1}-\epsilon)/\epsilon$. When the vacuum or oxide is absent $d=0$, we recover the NM/FMM/substrate trilayer reported in Ref.~20, where $G_{r,i}$ shows oscillatory behavior with respect to both the FMM thickness and the $s-d$ coupling, which has been attributed to the quantum interference effect when the spin travels inside the FMM. In the presence of the vacuum or oxide layer $d\neq 0$, the quantum interference effect still manifests, but the interference pattern changes significantly with the vacuum thickness $d$. Thus the vacuum or oxide layer, despite nonmagnetic, participates in and significantly influences the quantum interference effect for the spin transport in this multilayer system.

\section{Detail of linear response theory and differential conductance}

In NM/vacuum/FMI, the STT is equivalent to the torque produced by the effective magnetic field described by Eq.~(10). We may write the $z$-component of the torque operator as
\begin{eqnarray}
&&\hat{\tau}^{z}=\tilde{G}_{i}\left[{\bf S}_{0}\times{\boldsymbol\mu}_{0}\right]^{z}+\tilde{G}_{r}\left[{\bf S}_{0}\times\left({\hat{\bf S}}_{0}\times{\boldsymbol\mu}_{0}\right)\right]^{z}
\nonumber \\
&&=\tilde{G}_{i}\left[S_{0}^{x}\mu_{0}^{y}-S_{0}^{y}\mu_{0}^{x}\right]
\nonumber \\
&&+\tilde{G}_{r}\left[S_{0}^{x}\left({\hat S}_{0}^{z}\mu_{0}^{x}-{\hat S}_{0}^{x}\mu_{0}^{z}\right)-S_{0}^{y}\left({\hat S}_{0}^{y}\mu_{0}^{z}-{\hat S}_{0}^{z}\mu_{0}^{y}\right)\right]
\nonumber \\
&&=\frac{1}{2}\left(\tilde{G}_{r}\cos\theta+i\tilde{G}_{i}\right)S_{0}^{+}\mu_{0}^{-}
+\frac{1}{2}\left(\tilde{G}_{r}\cos\theta-i\tilde{G}_{i}\right)S_{0}^{-}\mu_{0}^{+}
\nonumber \\
&&\;\;-\frac{1}{2}\tilde{G}_{r}\sin\theta\left(e^{-i\varphi}S_{0}^{+}\mu_{0}^{z}+e^{i\varphi}S_{0}^{-}\mu_{0}^{z}\right)\;,
\label{tau_before_gauge}
\end{eqnarray}
where we have used the operators $S_{0}^{\pm}=S_{0}^{x}\pm iS_{0}^{y}$ and $\mu_{0}^{\pm}=\mu_{0}^{x}\pm i\mu_{0}^{y}$, and the unit vector ${\hat{\bf S}}_{0}=(\sin\theta\cos\varphi,\sin\theta\sin\varphi,\cos\theta)$. Likewise, the effective interacting Hamiltonian that gives ${\hat{\boldsymbol\tau}}=i\left[H_{int},{\bf S}_{0}\right]$ is
\begin{eqnarray}
&&H_{int}=-\tilde{G}_{i}{\bf S}_{0}\cdot{\boldsymbol\mu}_{0}-\tilde{G}_{r}{\bf S}_{0}\cdot\left({\hat{\bf S}}_{0}\times{\boldsymbol\mu}_{0}\right)
\nonumber \\
&&=-\tilde{G}_{i}\mu_{0}^{z}S_{0}^{z}-\frac{\tilde{G}_{i}}{2}\left(\mu_{0}^{+}S_{0}^{-}+\mu_{0}^{-}S_{0}^{+}\right)
\nonumber \\
&&-\frac{i}{2}\tilde{G}_{r}\cos\theta\left(\mu_{0}^{+}S_{0}^{-}-\mu_{0}^{-}S_{0}^{+}\right)
\nonumber \\
&&-\frac{i}{2}\tilde{G}_{r}\sin\theta\left(-e^{-i\varphi}\mu_{0}^{+}S_{0}^{z}+e^{i\varphi}\mu_{0}^{-}S_{0}^{z}\right.
\nonumber \\
&&\left.+e^{-i\varphi}\mu_{0}^{z}S_{0}^{+}-e^{i\varphi}\mu_{0}^{z}S_{0}^{-}\right)\;.
\nonumber \\
\label{Hint_before_gauge}
\end{eqnarray}
The $e^{\pm i\varphi}$ in Eqs.~(\ref{tau_before_gauge}) and (\ref{Hint_before_gauge}) can be removed by a gauge transformation $e^{\mp i\varphi}S_{0}^{\pm}\rightarrow S_{0}^{\pm}$, $e^{\mp i\varphi}\mu_{0}^{\pm}\rightarrow \mu_{0}^{\pm}$. After Fourier transform, one obtains Eq.~(11).

The linear response theory in Eq.~(12) leads to the calculation of 
\begin{eqnarray}
&&\tau^{z}=-i\int_{-\infty}^{\infty}dt^{\prime}\theta(t-t^{\prime})\langle\left[{\hat\tau}^{z}(t),H_{int}^{rel}(t^{\prime})\right]\rangle
\nonumber \\
&&=\int_{-\infty}^{\infty}dt^{\prime}\theta(t-t^{\prime})
\left\{-|J_{1}|^{2}e^{-i\mu_{0}(t-t^{\prime})}\langle\left[A^{\dag}(t),A(t^{\prime})\right]\rangle\right.
\nonumber \\
&&\;\;\;\;\;\;\;+|J_{1}|^{2}e^{i\mu_{0}(t-t^{\prime})}\langle\left[A(t),A^{\dag}(t^{\prime})\right]\rangle
\nonumber \\
&&\left.\;\;\;\;\;\;\;-|J_{2}|^{2}\langle\left[B^{\dag}(t),B(t^{\prime})\right]\rangle+|J_{2}|^{2}\langle\left[B(t),B^{\dag}(t^{\prime})\right]\rangle\right\}\;,
\nonumber \\
\end{eqnarray}
which requires the calculation of the Matsubara response function in Eq.~(14). The $L$ and $M$ channel give
\begin{eqnarray}
&&U^{L}(i\omega)=-\sum_{kk^{\prime}q}\int_{0}^{\beta}d\tau e^{i\omega\tau}\chi^{-+}(q,\tau)G_{\uparrow}(k,-\tau)G_{\downarrow}(k^{\prime},\tau)\;,
\nonumber \\
&&U^{M}(i\omega)=-\sum_{kk^{\prime}q}\int_{0}^{\beta}d\tau e^{i\omega\tau}\chi^{-+}(q,\tau)
\nonumber \\
&&\;\;\;\times\left[G_{\uparrow}(k,-\tau)G_{\uparrow}(k^{\prime},\tau)
+G_{\downarrow}(k,-\tau)G_{\downarrow}(k^{\prime},\tau)\right]\;,
\nonumber \\
\end{eqnarray}
where $G_{\sigma}(k,\tau)$ and $\chi^{-+}(q,\tau)$ are electron and magnon Green's function
\begin{eqnarray}
G_{\sigma}(k,\tau)&=&-\langle T_{\tau}c_{k\sigma}(\tau)c_{k\sigma}^{\dag}(0)\rangle=\frac{1}{\beta}\sum_{\omega_{n}}\frac{e^{-i\omega_{n}\tau}}{i\omega_{n}-\xi_{k\sigma}}\;,
\nonumber \\
\chi^{-+}(q,\tau)&=&-\langle T_{\tau}S_{-q}^{-}(\tau)S_{q}^{+}(0)\rangle=\frac{-2\langle S^{z}\rangle}{\beta}\sum_{\nu_{m}}\frac{e^{-i\nu_{m}\tau}}{i\nu_{m}+\omega_{q}}\;,
\nonumber \\
\end{eqnarray}
with $\beta=1/k_{B}T$. The frequency sum leads to 
\begin{widetext}
\begin{eqnarray}
U^{L}(i\omega)&=&2\langle S^{z}\rangle\sum_{kk^{\prime}q}\frac{\left[n_{F}(\xi_{k^{\prime}\downarrow})-n_{F}(\xi_{k\uparrow})\right]\left[n_{B}(\xi_{k\uparrow}-\xi_{k^{\prime}\downarrow})-n_{B}(-\omega_{q})\right]}
{i\omega+\xi_{k\uparrow}-\xi_{k^{\prime}\downarrow}+\omega_{q}}
\nonumber \\
U^{M}(i\omega)&=&2\langle S^{z}\rangle\sum_{kk^{\prime}q}\frac{\left[n_{F}(\xi_{k^{\prime}\uparrow})-n_{F}(\xi_{k\uparrow})\right]\left[n_{B}(\xi_{k\uparrow}-\xi_{k^{\prime}\uparrow})-n_{B}(-\omega_{q})\right]}
{i\omega+\xi_{k\uparrow}-\xi_{k^{\prime}\uparrow}+\omega_{q}}
\nonumber \\
&+&2\langle S^{z}\rangle\sum_{kk^{\prime}q}\frac{\left[n_{F}(\xi_{k^{\prime}\downarrow})-n_{F}(\xi_{k\downarrow})\right]\left[n_{B}(\xi_{k\downarrow}-\xi_{k^{\prime}\downarrow})-n_{B}(-\omega_{q})\right]}
{i\omega+\xi_{k\downarrow}-\xi_{k^{\prime}\downarrow}+\omega_{q}}\;.
\nonumber \\
\end{eqnarray}
\end{widetext}
Using analytical continuation and ${\rm Im}\left[\frac{1}{x+i\eta}\right]=-\pi\delta(x)$, one obtains Eq.~(15), in which the $U^{M}(i\omega)$ response function does not contribute.

To see how the derivative of spin current is related to magnon density of states (DOS), consider the integral in Eq.~(15) denoted by $\tilde{I}$. We make the usual assumption in tunneling spectroscopy that the DOS of the probe (the NM) stays constant within the range of measurement so it can be pulled out of the integration, and notice that the $n_{B}(-\Omega)$ factor does not contribute to the integral, thus
\begin{widetext}
\begin{eqnarray}
\tilde{I}&=&N_{\uparrow}N_{\downarrow}\int_{-\infty}^{\infty}d\xi\int_{0}^{-\mu_{0}}d\Omega
\left[n_{F}(\xi+\mu_{0}+\Omega)-n_{F}(\xi)\right]n_{B}(-\Omega-\mu_{0})N_{M}(\Omega)
\nonumber \\
&\approx&N_{\uparrow}N_{\downarrow}\int_{-\xi_{0}}^{\xi_{0}}d\xi\int_{0}^{-\mu_{0}}d\Omega
\left[\frac{1}{4}+\frac{(\Omega+\mu_{0})}{8k_{B}T}-\frac{\xi^{2}}{16k_{B}^{2}T^{2}}
-\frac{\xi(\Omega+\mu_{0})}{16k_{B}^{2}T^{2}}\right]N_{M}(\Omega)\;,
\nonumber \\
\label{I_integral_approximation}
\end{eqnarray}
\end{widetext}
where we expand the distribution function to leading order in $\xi$ and $\Omega+\mu_{0}$, and notice that $-\mu_{0}>0$. The distribution function is mainly concentrated within the range of Fermionic energy $-\xi_{0}<\xi<\xi_{0}$ whose boundary is proportional to temperature $\xi_{0}\propto k_{B}T$ (using the expansion in Eq.~(\ref{I_integral_approximation}) and solve for where the distribution function vanishes at $\Omega=-\mu_{0}$ yields $\xi_{0}=2k_{B}T$). Differentiating the integral with respect to $-\mu_{0}$ gives, using either numerical integration or the approximated analytical form of Eq.~(\ref{I_integral_approximation}),
\begin{eqnarray}
\frac{d\tilde{I}}{d(-\mu_{0})}\propto N_{\uparrow}N_{\downarrow}N_{M}(-\mu_{0})k_{B}T\;,
\end{eqnarray}
and so follows Eq.~(16). We anticipate that the linear $T$ dependence may help to separate the calculated tunneling spin conductance from other contributions to the spin current if they have a different power law dependence on temperature.

\end{document}